\documentclass[letterpaper,twocolumn,10pt]{article}
\usepackage{secure-email-transmission}

\usepackage{tikz}
\usepackage{amsmath}

\usepackage{filecontents}

\begin{document}
%-------------------------------------------------------------------------------

%don't want date printed
\date{}

\title{\Large \bf  Secure Email Transmission Protocols \\
	— A New Architecture Design}

%for single author (just remove % characters)
\author{
{\rm Gabriel Chen}\\
University of Illinois Urbana Champaign \\
gc30@illinois.edu
\and
{\rm Rick Wanner}\\
SANS Institute \\
rwanner@pobox.com
% copy the following lines to add more authors
% \and
% {\rm Name}\\
%Name Institution
} % end author

\maketitle

%-------------------------------------------------------------------------------
\begin{abstract}
%-------------------------------------------------------------------------------
During today’s digital age, emails have become a crucial part of communications for both personal and enterprise usage. However, email transmission protocols were not designed with security in mind, and this has always been a challenge while trying to make email transmission more secure. On top of the basic layer of SMTP, POP3, and IMAP protocols to send and retrieve emails, there are several other major security protocols used in current days to secure email transmission such as TLS/SSL, STARTTLS, and PGP/GPG encryption. The most general design used in email transmission architecture is SMTP with PGP/GPG encryption sending through an TLS/SSL secure channel. Regardless, vulnerabilities within these security protocols and encryption methods, there is still work can be done regarding the architecture design. In this paper, we discuss the challenges among current email transmission security protocols and architectures. We explore some new techniques and propose a new email transmission architecture using EEKS structure and Schnorr Signature to eliminate the usage of PGP/GPG for encryption while achieving Perfect Forward Secrecy.

\end{abstract}

%-------------------------------------------------------------------------------
\section{Introduction}
%-------------------------------------------------------------------------------

Email has become an integral part of commerce and personal daily life. Since the time emails started being used to exchange important information and treated as written evidence, email security, or digital information transmission security in general, has always been a challenge within both personal and enterprise usage. Without adding too much complexity onto the transmission architecture, a secure yet simple email transmission design is desired. Considering the original email transmission protocol without security built in, SMTP, has been deployed over all these years towards general email servers, redesigning from scratch with security in mind is not feasible. Through years, researchers and engineers are building encryption and compression blocks on top of SMTP to make email transmission more secure \cite{Bhatt2019EMailSF,  Jain2008EmailSU, Kozachok2021ModelOP,  Om2017SecureEG}, and some of the research has a specific system target such as Android system \cite{Kisembo2021AnAF}. Now, the transmission process is indeed much more secure than the original one used when the first email was sent. However, there are still many weakness and vulnerabilities in current email transmission architecture design. 

Here, we present a new architecture design using EEKS architecture with Schnorr Signature to secure email transmission with Perfect Forward Secrecy. In Section 2, we go over the fundamental email transmission architecture design with an introduction to basic email transmission protocols. In Section 3, we discuss vulnerabilities in our current general email transmission architecture designs and show how urgent it is for a better architecture. In Section 4, we discuss some modern approaches to a better email transmission architecture design including EEKS architecture, DMARCBox, quantum entangling for authentication, and Schnorr Signature. Finally, we discuss our new architecture which combines the power of EEKS and Schnorr Signature in Section 5. 

%-------------------------------------------------------------------------------
\section{Preliminary}
%-------------------------------------------------------------------------------

To have a better grasping of the current email architecture design, below we give a refresher regarding definitions of some fundamental protocols and terminologies utilized while sending and receiving emails.

\subsection{Fundamental Email Protocols and Terminologies}
%-----------------------------------

\begin{description}
\item[SMTP] SMTP stands for Simple Mail Transfer Protocol \cite{emailprotocolstutorialspoint, walsh_2020}. This is the most standard email transmission protocol used to send emails. It is transferred in pain text without any security or encryption methods built on top.  However, it is reliable to guarantee the transmission as it will provide an error message back to the sender if the email failed to be delivered.

\item[POP] POP stands for Post Office Protocol, and POP3 is POP version 3 \cite{Orman2015EncryptedET}. It is utilized while downloading emails from email servers. When POP is used, emails are automatically deleted after first retrieve. This makes POP good for one single client accessing one single mailbox with less internet usage and more independency from the server. 

\item[IMAP] IMAP stands for Internet Message Access Protocol \cite{Orman2015EncryptedET}. It is also an email receiving protocol. It is different from POP as IMAP does not delete messages after retrieving. Thus, IMAP can support multiple clients accessing one single mailbox. However, IMAP will consume more network usage and is more dependent on the server. 

\item[SSL/TLS] TLS stands for Transport Layer Security and SSL stands for Secure Socket Layer \cite{walsh_2020}. After a TCP channel is established, a SSL/TLS certificate will be sent and a secure channel will be established after the SSL/TLS handshake.\cite{smtptls} Today, TLS is more secure than SSL and should be a replacement of SSL \cite{kinsta_2021}. 

\item[STARTTLS] STARTTLS is the protocol which attempts to establish a secure SSL/TLS channel \cite{griffin_2020}. It is used with SMTP, IMAP, and POP3. The STARTTLS process happens before the SSL/TLS handshake and it will either establish a secure SSL/TLS tunnel or fall back to simple SMTP depends on the design. If the design has enforced TLS, the fallback strategy is ending the connection; otherwise if the design is opportunistic TLS, the fallback strategy is transmitting without TLS.

\item[PGP/GPG]  PGP stands for Pretty Good Privacy and GPG stands for GNU Privacy Guard. GPG is a free alternative to the PGP program \cite{differencebetweenpgpandgpg2021}. PGP is an asymmetric key encryption program which utilize cryptography to support data confidentiality and authentication. PGP utilizes RSA and IDEA for encryption, and GPG utilizes NIST and AES as its encryption algorithms. 

\item[SPF] SPF stands for Sender Policy Framework. It is an authentication protocol to prevent sender’s address forgery \cite{Orman2015EncryptedET}. Since SMTP does not validate the sender’s address, anyone can potentially put whatever one wants in the sender’s address. To mitigate this vulnerability, a type of DNS record contains SPF header which has a list of domains which are authorized to send emails. When an email send attempt is made, sender’s domain is verified using this DNS record to detects any unauthorized activities. If any unauthorized email sending is detected, the email is rejected. 

\item[DKIM] DKIM stands for Domain Keys Identified Mail \cite{Orman2015EncryptedET}. It also mitigates the sender’s address forgery. DKIM uses digital signature with asymmetric keys to verify the sender is who they claim to be, and it also verifies the content of the email is not changed. DKIM is more advanced than SPF as it solves the issue that SPF allows same server forgery.

\item[DMARC] DMARC stands for Domain Based Message Authentication Reporting and Conformance \cite{Orman2015EncryptedET}. It is another email authentication protocol which works with DKIM and SPF. These three protocols are usually used together to mitigate email sender forgery and strengthen the overall security.

\item[S/MIME] S/MIME stands for Secure/Multipurpose Internet Mail Extensions \cite{walsh_2020}. It is a protocol to support end-to-end email encryption and support by most email servers. S/MIME uses digital certificates to guarantee the authentication and confidentiality of emails, and it encrypts of the contents of emails before sending them out to the internet to achieve end-to-end encryption. The downside of S/MIME is that it only encrypts the contents but not the header or metadata.
\end{description}
	
\subsection{Fundamental Design of Email Transmission}
%-----------------------------------

In current general email transmission designs, emails are sent through SMTP, and received using POP3 / IMAP \cite{jithin_2016, sparkpost_2018}. However, pure SMTP and POP3/IMAP are transmitted using plaintext which do not provide any security. To add security, these protocols are encapsulated in a secure transmission channel such as SSL/TLS spanned through STARTTLS. In this design, the transmission tunnel is secured but not the message itself. If the channel is compromised, the message can be read in plaintext. Therefore, encryption algorithms such as GPG/PGP are used to encrypt messages going through SSL/TLS tunnels to further strength the security.

With the knowledge of above fundamental email transmission protocols, we will discuss the basic fundamental email transmission design used in modern days. The general design include two clients sides and two email servers sides. Let us assume Alice and Bob are trying to send emails to each other, and Alice initiate the first ever email. Alice’s email is constructed on Alice’s client’s side, then transmitted to Alice’s email server which also called a Mail Transfer Agent (MTA). Next, Alice’s email will go from Alice’s email server to Bob’s email server, and it will be sent to Bob’s client side eventually.

%---------------------------
\begin{figure}
	\begin{center}
		\includegraphics[width=\linewidth]{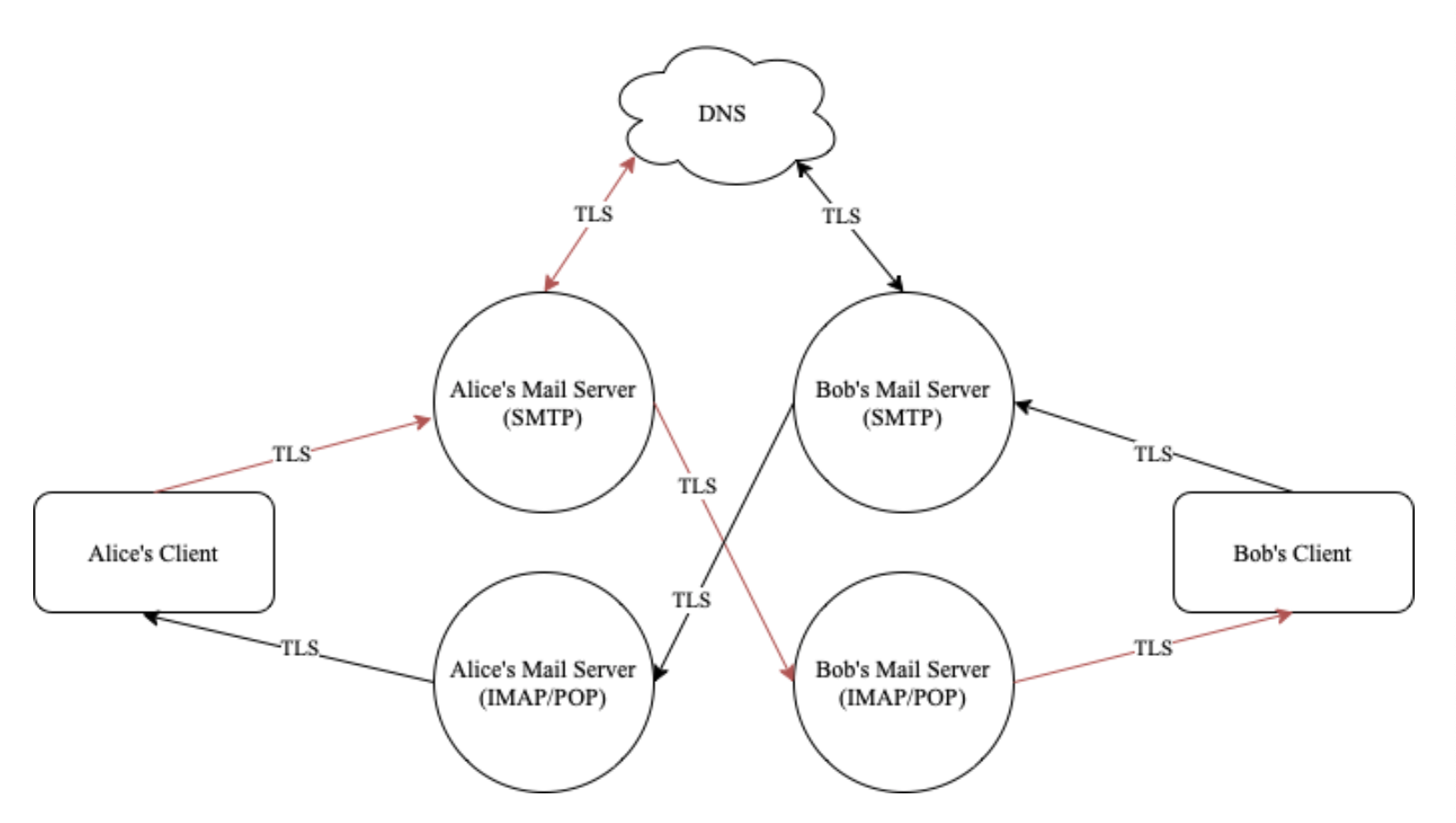}
	\end{center}
	\caption{\label{fig:email} Email Transfer from Alice to Bob}
\end{figure}

As we see in figure 1, after Alice clicks the “send” button on her client side, her email can be encrypted using either PGP/GPG or S/MIME. In PGP/GPG, either RSA or AES encryption algorithm is used. After the encryption, the cipher text is transferred to Alice’s MTA using SMTP within a secure TLS channel after the STARTTLS attempt. Then, a DNS look up will be performed to find Bob’s MTA. After getting the address of Bob’s MTA, Alice’s email will be transferred to Bob’s MTA also through SMTP. If Alice’s email server and Bob’s email server are in different domains, different security design might involve, and there is a fall back or connection drop choice for Alice’s MTA, which means if Bob’s MTA does not support security requirements which are enforced by Alice’s MTA, the connection might drop. For example, Bob’s email server does not support TLS, then if Alice’s email server has enforced TLS, the connection will be dropped; however, if Alice’s email server has opportunistic TLS, the connection will fall back to a non TLS channel. When the email successfully arrives at Bob’s MTA and been verified using SPF, DKIM and DMARC, it will be retrieved by Bob’s client side using either IMAP or POP3. Both IMAP and POP3 email retrieving protocols are also wrapped around with a secure TLS channel. 

%-------------------------------------------------------------------------------
\section{Vulnerabilities in Current Designs}
%-------------------------------------------------------------------------------

Although it looks like we already have a good design which is able to achieve secure email exchange, there are still vulnerabilities and limitations within our transmission channel, encryption methods, and the design itself.
 
\subsection{Vulnerabilities in the Transmission Channel}
%-----------------------------------

Regardless of the overall design, there are vulnerabilities in the blocks we are using to build the whole email transmission process. In the current design, TLS is utilized as the protocol to establish a secure channel to transmit emails on top of SMTP, POP, or IMAP. To start a TLS secure channel, a TLS handshake is initiated first to agree on a cipher, a master secret, and certificates between two sides of the communication. The TLS agreement looks like the following:

$$\texttt{TLS\_DHE\_RSA\_WITH\_AES\_256\_CBC\_SHA256}$$

Observe that the above TLS agreement uses RSA encryption with an AES 256 encryption method and CBC mode. The integrity is enforced using the SHA 256 hashing algorithm. As we know, to establish a TLS secure channel, STARTTLS will first attempt and see whether the other side of the communication supports TLS. Although under modern designs, TLS is enforced in most servers and STARTTLS is not necessary in most cases, many architects still implement STARTTLS before attempting the TLS handshake to make sure the connection can be established. After all, we care about the availability of the email transmission service more than other properties in CIA triad.

%---------------------------
\begin{figure}
	\begin{center}
			\includegraphics[width=\linewidth]{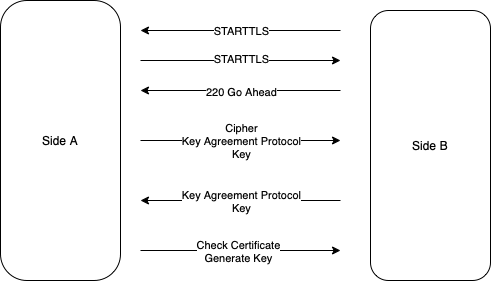}
\end{center}
\caption{\label{fig:starttls} STARTTLS and TLS handshake}
\end{figure}

However, both the STARTTLS protocol and the TLS handshake are vulnerable to Machine in the Middle (MitM) attack the same as many other handshake processes. When Side A attempts its first STARTTLS message, it is unencrypted. the attacker could sit in between Side A and Side B then perform a TLS Stripping attack. To improve availability, STARTTLS for email may use opportunistic TLS which falls back to a non-secure channel when TLS is not support on the other side. In TLS stripping \cite{ssltlsvulnerabilitie}, the attacker impersonates Side B and tells Side A that TLS is not support, then Side A will send out message in an unsecured channel. In the end, the attacker is able to read everything being transmitted from Side A to Side B. 

Besides stripping attack, TLS is also vulnerable to certificates attack as described in \cite{costlow_2021}. Wildcard certificates are used by many system administrators to save time and effort while creating certificates for each domain. For example, a wildcard certificate “*.example.com” is valid through “www.example.com”, “ftp.example.com”, “smtp.example.com”, and etc. If this certificate is compromised which means the attacker gained the private key of this certificate, the attacker is able to decrypt any message transferred through TLS using this certificate. On the other hand, if the server only utilize Certificate Authorities (CA) signed certificate, there is a still possibility that the CA is compromised or forged by attackers.

\subsection{Vulnerabilities in Encryption Methods}
%-----------------------------------

Besides the vulnerabilities that lie within the TLS secured transmission channel, there are also vulnerabilities in encryption methods used in email exchange. In email transmission, PGP/GPG and S/MIME are two major encryption methods to encrypt the message itself while data is at rest. 

\subsubsection{GPG CVE-2021-3345 Overflow Vulnerability}

GPG is one of the most used encryption methods among email encryption at rest. Major email clients such as Thunderbird and Apple Mail use GPG as their encryption method to make emails more secure within the TLS channel. This is one of the reasons that GPG vulnerabilities are very severe based on a huge user amount. As users tend to believe something under a massive usage has to be flawless, it adds on to the influence of a GPG vulnerability.

In GPG Libgcrypt version 1.9.0, there is a heap-based buffer overflow in file \texttt{\_gcry\_md\_block\_write}. When the digest final function sets a large count value, this vulnerability, CVE-2021-3345 \cite{cve-2021-3345}, is able to compromise confidentiality, integrity, and availability of the message being encrypted using GPG. It is recommended to update to version 1.9.1 or later, but this vulnerability is still under investigation without knowing how much damage it has caused until it was disclosed in year 2021.

\subsubsection{S/MIME and CVE-2020-1774}

As we have addressed in previous section, S/MIME does not encrypt the email header nor the metadata, and there are many different S/MIME implementation standards among many different servers which makes it hard to scale. Thus, there comes a fall back strategy that if the recipient’s  S/MIME standard does not support the sender’s S/MIME standard, it cannot decrypt the message encrypted under sender’s S/MIME standards.

Besides some draw back S/MIME naturally has, there are also vulnerabilities among S/MIME practice in real life. In CVE-2020-1774 \cite{cve-2020-1774}, when user downloads S/MIME keys or certificates, the name for private and public keys are the same in exported file. This gives the possibility to mix public keys and private keys, then send the private key to the other side of communication instead of the public key. As we can see, this vulnerability breaks the confidentiality while transmitting emails, and it requires very few exploitation skills for attackers to exploit it.

\subsubsection{Vulnerabilities in the Design}

Considering the lack of a scalable secure email transmission design, the most general email transmission process needs to consider a lot of different situation and edge cases to make the decision to either fall back or end the connection. Unlike other internet protocols, email transmission requires much more availability than others in the CIA triad. As the strong requirement towards availability, many designs intend to fall back to non-secure protocols instead of ending connections all together. This increases the impact of a vulnerability where a client is under an old and/or insecure mail server which does not support the most advanced security protocols in modern designs. 

%-------------------------------------------------------------------------------
\section{New Design Approaches to Achieve Security}
%-------------------------------------------------------------------------------

Based on the awareness that the fundamental email transmission architecture is not secure, there are several approaches which can be made to design a better email exchange architecture. In general, there are two major approaches where we can attack this problem. We can either enhance our current fundamental design, or introduce emerging technologies which have evolved dramatically through recent years.

\subsection{Current Structure Enhancement}
%-----------------------------------

As most email exchange architectures and protocols are already implemented, the most achievable solution to make them more secure is to add enhancements on top of what we already have without redesigning the whole structure. In current well developed protocols such as SMTP, PGP/GPG, and DMARC, there is still room for security enhancement.

\subsubsection{EEKS Architecture}

Enhanced Encryption Key System, or EEKS, is a new architecture proposed to enhancement the security of SMTP using asymmetric cryptography \cite{Nourai2017SecuringEF}. The goal of EEKS is to massively improve the accessibility to email encryption for the general public by eliminating the usability issues of PGP/GPG. 

In order to eliminate PGP/GPG but retain the asymmetric cryptography, the process of public key exchange has to be restructured. In EEKS, it incorporates asymmetric cryptography mechanism directly into SMTP. It extends the SMTP command set to include an ENCRYPT command that provides cryptography and hash functions, FINGERPRINT contains the digital signature of the original message, TIME2LIVE enables the ability of self-destruction, and SESSION serves as a serial of random numbers which is utilized as salt. After the SMTP gains cryptography power, EEKS also introduces a local EEKS server and a root EEKS server to store the key pair to replace the functionality of CA in PGP/GPG. Essentially, local EEKS residents on the user LAN and root EEKS server lives on the Internet where local EEKS will periodically check in with root EEKS server for public key creation and revocation. 

There are several benefits of an EEKS architecture. First of all, the elimination of GPG/PGP service reduces the development difficulty while building an email server as the enhanced SMTP under EEKS carries encryption by itself. Secondly, the separation of User Agent (UA), EEKS local server, and EEKS gives additional protection against the compromise of email account. Thirdly, the automatic asymmetric cryptography key generation process enhances the message confidentiality, integrity, and non-repudiation. 

However, like any other security architecture, EEKS has its own weakness. For example, the connection between all EEKS servers are protected by TLS/SSL which has its vulnerabilities as we discussed above. Also, the EEKS encryption process can adds additional difficulty in searching emails.

%---------------------------
\begin{figure}
	\begin{center}
		\includegraphics[width=\linewidth]{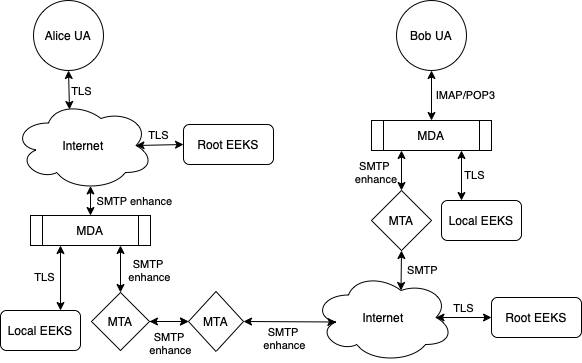}
	\end{center}
	\caption{\label{fig:eeks} EEKS Architecture }
\end{figure}

\subsubsection{DMARCBox}

Besides email message encryption, email spoofing prevention has been a major email security topic for years. As there is no spoofing prevention built into SMTP, SPF and DKIM were developed. On top of SPF and DKIM, DMARC was built to better prevent sender spoofing. 

However, in order to have a better prevention mechanism, we have to have a better detection method. To detect and react as early as possible, a better analysis tool is needed to identify malicious emails. Currently, the most popular technique is the spam filter logic. Techniques include but not limited to Classification, Categorization, Profiling (Machine Learning), Natural Language Processing, Parsing and Filtering when spam filter algorithms are built. The vulnerability for a spam folder is that emails are analyzed after they are already delivered, and it significantly increase the possibility where a user can easily open the spam folder and phished by malicious emails. To build a better mechanism for malicious email detection, we need to analyze emails during transmission and react before they are in the mailbox. 

\begin{figure}
	\begin{center}
		\includegraphics[width=\linewidth]{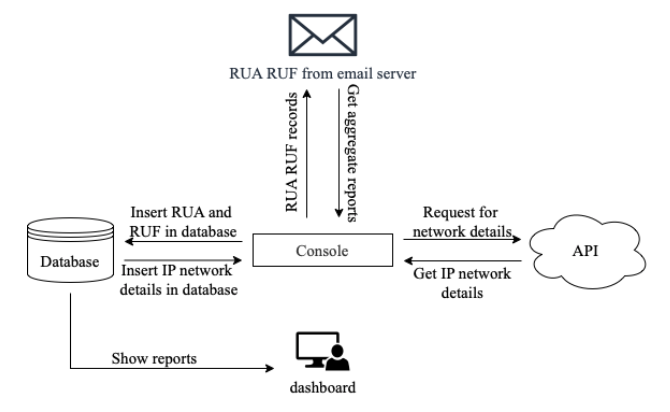}
	\end{center}
	\caption{\label{fig:dmarc} DMARCBox Workflowe }
\end{figure}

DMARCBox \cite{Nanaware2019DMARCBoxC} is built for the purpose of analyzing malicious emails during transmission. It utilize the DMARC protocol to retrieve the Reporting URIs for Aggregate Date (RUA) and Reporting URIs for Failure (RUF) records, then provides highly readable reports based on the analysis of the email. It can be developed as either onsite or cloud based. The overall workflow of DMARCBox can be found in Figure 4. 

Since DMARC is built on top of SPF and DKIM, DMARCBox can also utilize the protocol details of SPF and DKIM while classifying emails. DMARCBox provides seven different categories according to RUA and RUF, and additional reaction protocols can be built according to these classifications.

\subsection{Introduce New Technologies}
%-----------------------------------

As new technologies have been evolving dramatically since the first appearance of SMTP, many researchers start considering to incorporate them into their new design approaches to build a secure email architectures. New technologies such as quantum teleportation and heavy mathematical approaches, come in to play.  

\subsubsection{Quantum Teleportation}

As we move further with email transmission security, new technologies are eagerly expected to be introduced as we are getting more and more vulnerabilities within current protocols every single day.  Since the majority of email services grant their users enough power to perform any activities they would like once users are authenticated, the process of authentication is crucial. To achieve faster and more secure email authentication, much research has been conducted such as \cite{Nemavarkar2015AUA}. To untangle the issue of a strong user authentication process, Quantum Teleportation \cite{Shufen2016ResearchOA} as a new technology is introduced. 

Instead of relying on the algorithmic complexity in Public Key Infrastructure (PKI), quantum teleportation achieves authentication and encryption based on non-locality of entanglement correlation space. Before the communication, both sides’ Mail Server (MS) will be authenticated by the Authentication Center (CA). Then, they will use a shared quantum encryption key for email encryption through communication. The overall authentication process is shown as in Figure 5.

\begin{figure}
	\begin{center}
		\includegraphics[width=\linewidth]{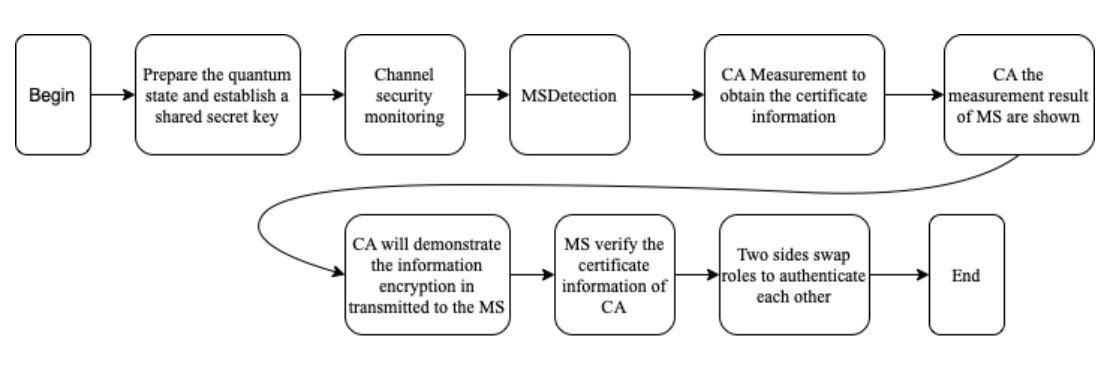}
	\end{center}
	\caption{\label{fig:quant} Quantum Authentication }
\end{figure}

In general, this process takes five steps to achieve users quantum authentication. In step 1, quantum entangled states are prepared, and a shared secret key is established through the quantum distribution protocol BB84. In step 2, the communication channel established through step 1 is closely monitored by selecting photons by both CA and MS. If both CA and MS generates the same measurement results from these photons, the channel is secure and safe. In step 3, MS will measure a special quantum state based on MB with a result for step 4. In step 4, CA will verify the measurement results from MS in step 3. If there is anything leads to a fake MS identifier, CA will deny the result and the channel needs to be ended because of compromise. Otherwise, step 5 will carry on. In step 5, CA will gather the demonstration information based on quantum encoding ready for next step. In step 6, CA will encrypt the demonstration information from step 5 and send it to MS using the shared secret key they exchanged in step 1. In step 7, MS receives the information from CA sent in step 6 and verifies it by comparing it with its own demonstration information. In step 8, two parties will exchange their roles and verify each other again. To elaborate, MS will gather its demonstration information and encrypt it using the shared key, then send it to CA for verification. CA will decrypt the message and verify it with its own demonstration information. Up until now, the authentication process has finished and both CA’s and MS’ identities are verified. 

Although the quantum teleportation authentication process utilizes the quantum space to significantly reduce some security risks we are facing right now in our current structures such as vulnerabilities in GPG/PGP and TLS/SSL, it has its own drawbacks. The major drawback of this design is the quantum teleportation is not yet achievable by the general public. Besides, this protocol will still suffer attacks against quantum space and quantum technologies once it is massively achievable.

\subsubsection{Schnorr Signature}

One of drawbacks in current email architecture design lies within the incorporation of encryption mechanism. Regardless the usage of either asymmetric or symmetric encryption, once the master encryption key is compromised, all previous emails which were encrypted using the same key are readable. Thus, lots of research is done to design email protocols with Perfect Forward Secrecy (PFS) against MITM (Machine in the Middle) attack. 

Here, we discuss an email protocol with PFS addressed in \cite{2021Symm...13.1144L}. It incorporates Schnorr signature with a registration phase for both sending phase and receiving phase. We assume the verification service for registration is trust worthy. This verification process can be built into our email servers. Before each sending and receiving action is performed, an extra registration phase is performed to verify both sender and receiver. After the registration, a session key is generated based on Schnorr signature for both sending and receiving activities. Thus, even the master encryption key which encrypts emails contents is compromised, the session key is an extra layer to provide PFS against eavesdropping.

Below is the process of generating a Schnorr signature pair $(e, s)$  to send a message \footnote{h is a one way has function, and || is the concatenation operator.}. 

1. Compute $r \equiv g^\delta mod p$, where $\delta$ is a randomly chosen integer $1 \leq \delta < q$.

2. Compute $e \equiv h (M || r)$.

3. Solve $s \equiv xe + \delta mod q$.

And the verification of Schnorr signature is the following process.

1. Obtain the signer public key $(p, q, g, y)$.

2. Compute $r' \equiv g^sy^{-e} mod p$.

3. Compute $e' \equiv h (M || r')$.

4. Accept the signature if and only if $e' = e$.

To break the session key, the attacker has to break the Schnorr signature. The security of Schnorr signature is guaranteed by the Discrete Logarithm Assumption among all discrete logarithm problems ($DLP r' \equiv g^s y^{-e} mod p$). In other word, we have to be able to easily find prime factors among all numbers to break the Schnorr signature. Under the general assumption that $P \neq NP$ in Computer Science, there does not exist such feasible algorithm to achieve it and will never be.

Although $P \neq NP$ assumption is generally assumed as true in Computer Science, it does not necessarily mean this assumption is absolutely true. we hold a neutral opinion towards this issue in this paper. If it is proved that $P = NP$, Schnorr Signature will become useless towards email transmission security. But after all, the whole Computer Science foundation will collapse in a sense once we prove $P = NP$, and Schnorr Signature will be one of the least thing we would worry about at that point.

%-------------------------------------------------------------------------------
\section{A New Design}
%-------------------------------------------------------------------------------

Below we propose a new design which combines the power of EEKS architecture and Schnorr Signature.

\subsection{Design Workflow}

In current EEKS architecture design, it gives us the opportunity to eliminate the usage of GPG/PGP in email transmissions and build our encryption mechanism directly into the SMTP protocol. However, one of the major issues lie within the EEKS architecture is that it does not provide any Perfect Forward Secrecy (PFS). If the master encryption key is compromised, all previous messages which were encrypted using this key can be decrypted and read. To mitigate this weakness, we can build Schnorr signature for all sessions into EEKS structure. In this Defense in Depth set up, we utilize Schnorr signature to generate another encryption layer for each send/receive session. Under this design, even if our EEKS master encryption key is compromised, all previous sessions were still encrypted using the Schnorr signature session key which provides us the PFS.

\subsection{Drawback}

The major drawback of our design is the assumption of $P \neq NP$. As discussed in the previous section, if it is shown that $P = NP$ at some point in the future, the Schnorr Signature is useless to enforce the security in email transmission. Thus, our design will not survive under the possibility of $P = NP$.

Another drawback of our design has to do with the fallback procedures of EEKS architecture. Since the encryption process is built into SMTP protocols under EEKS design, there is a chance that the other side of email server does not support enhanced SMTP which causes issues while decrypting the message. This can only be resolved by implementing enhanced SMTP on both side of email servers to retain the power of EEKS design. 

\subsection{Further Possibilities}

Beyond our major design which utilizes both EEKS and Schnorr Signature, DMARCBox can be our analysis tool for filtering malicious emails while in transmission. It is also useful when we set up allow list and transmission rules on our email servers. Additionally towards users authentication, once the quantum field is achievable by the general public, quantum entangling can be an authentication method for our users’ login process to the email client side.

%-------------------------------------------------------------------------------
\section{Conclusion}
%-------------------------------------------------------------------------------

 Email transmission security has been and will continue to be a major topic in Network Security. To achieve a better security result, a good design is crucial. The EEKS architecture is proven to be one of the ways to achieve a better security comparing to our current general email transmission design using GPG/PGP. On top of EEKS, Perfect Forward Secrecy can be achieved through Schnorr Signature for session encryption. In this way, we achieve a better designed SMTP protocol with PFS.
 
After building out our email servers architecture, a tool to constantly monitor and tweak servers’ settings is important while filtering out and blocking any malicious emails. We explored the possibility of DMARCBox which stops malicious emails in transmission and believe it is a good tool to build into email servers to perform useful analysis during transmission processes.

%-------------------------------------------------------------------------------
\section*{Further Research}
%-------------------------------------------------------------------------------

With our proposed new design which combines the EEKS architecture with Schnorr Signature, we can eliminate the usage of GPG/PGP and achieve PFS. However, we barely touched the transmission tunnel encryption which is currently utilizing the power of TLS/SSL. There is still a lot of work that can be done in this part.

Essentially, our design introduces some level of complexity comparing to the general email transmission architecture design. As we all aware, complexity can cause network delay. Although low latency is not too important in email transmission as we care about availability more than others in the CIA triad, more work can still be done in low latency architecture design.

As for the authentication process, we majorly discussed one possibilities using quantum entangling. This is not yet generally achievable under current technology limitations. As there are many other authentication techniques can be explored and incorporate for user login on the client side, further research can be conducted in this aspect.

%-------------------------------------------------------------------------------
\section*{Acknowledgments}
%-------------------------------------------------------------------------------

We thank WiCyS (Women in Cybersecurity) for their generous support to this paper.  

\newpage

%-------------------------------------------------------------------------------
\bibliographystyle{plain}
\bibliography{\jobname}

%%%%%%%%%%%%%%%%%%%%%%%%%%%%%%%%%%%%%%%%%%%%%%%%%%%%%%%%%%%%%%%%%%%%%%%%%%%%%%%%
\end{document}